\documentclass[twocolumn,preprintnumbers,nofootinbib]{revtex4}
\usepackage{amssymb}
\usepackage{float}
\usepackage{amsmath}
\usepackage{graphicx}
\usepackage{color}

\begin{document}
\newcommand*{\PKU}{School of Physics, Peking University, Beijing 100871, China}\affiliation{\PKU}
\title{Multi-Sommerfeld enhancement in dark sector}
\author{Zhentao Zhang}\email{zhangzt@pku.edu.cn}\affiliation{\PKU}
\begin{abstract}
We study the multi-Sommerfeld enhancement in the case where $V(r)$ is composed of different kinds of potentials. We show that there are special properties of the multi-Sommerfeld enhancement. The physical content of the multi-Sommerfeld mechanism is carefully demonstrated. The multi-Sommerfeld enhancement might play a role in dark matter annihilation.
\end{abstract}
\maketitle

\section{Introduction}

The Sommerfeld enhancement is a fundamental effect in non-relativistic
quantum mechanics \cite{Sommerfeld}, which characterizes the enhancement of
the cross section due to an attractive potential (in the case of the repulsive potential, the cross section
is suppressed):
\begin{equation}\sigma  = S{\sigma_0}
\end{equation}
\noindent
where $\sigma$ is the actual cross section, ${\sigma _0}$ is the cross section without potential,
and S is the Sommerfeld enhancement factor. This factor is also known as the Sommerfeld-Sakharov factor \cite{Sakharov}.

As we know, an incident plane wave traveling in the positive direction of the
z-axis is
\begin{equation}{\psi _0}(r) = {e^{ikz}}
\end{equation}
The scattering of a central potential must produce the wave
function that behaves like an incident plane wave plus an outgoing spherical
wave for $r\to\infty$, in the form
\begin{equation}\psi({r,\theta }) \to {e^{ikz}} + f(\theta)\frac{{{e^{ikr}}}}{r} ~as~r \to \infty \end{equation}
\noindent
and the solution of the Schr\"{o}dinger equation has the form
\begin{equation}
\psi = \frac{1}{{2k}}\sum\limits_{l = 0}^\infty  {{i^l}}( {2l + 1}){e^{i{\delta _l}}}{P_l}( {\cos \theta}){R_{kl}}(r)
\end{equation}
where ${\delta _l}$ is the phase shift and the ${R_{kl}}(r)$ are radial functions satisfying the
equation
\begin{equation}( - \frac{{{\partial ^2}}}{{\partial {r^2}}} - \frac{2}{r}\frac{\partial }{{\partial r}} + 2mV(r) + \frac{{l( {l + 1})}}{{{r^2}}}){R_{kl}}(r) = {k^2}{R_{kl}}(r)\end{equation}
\noindent
where $m$ is the particle mass, $V(r)$ is the central potential.

If the interaction is point-like and takes place in the origin, the Sommerfeld enhancement factor due to the potential is
\begin{equation}S = \frac{{{{|{\psi(0)}|}^2}}}{{{{| {{\psi _0}(0)}|}^2}}} = {|{\psi (0)}|^2}\end{equation}
We can solve the Schr\"{o}dinger equation to obtain the boost factor.

The essence of the Sommerfeld effect can be easily understood: the wave function of the particle is strongly distorted by the appearance of a non-relativistic potential, the plane wave can not be the good approximation of the incident state, thus the real cross section could be much different from the one which is not affected by the potential.

The Sommerfeld enhancement is very important for us to understand the phenomenon of positron fraction excess in cosmic ray. This mechanism itself and its applications in dark matter annihilation were extensively studied \cite{Hisano,Hisano2,Cirelli,March-Russell,Strumia,March-Russell2,Arkani-Hamed,Lattanzi,Iengo,Cassel,Slatyer}.

However, in this paper, we will study the Sommerfeld effect in a different case where $V(r)$ is composed of different kinds of potentials. This situation is frequently encountered in practice. In contrast to other simple
potential cases, we will refer to this kind of effect as the multi-Sommerfeld enhancement.

\section{Multi-Sommerfeld enhancement}

The multi-Sommerfeld enhancement could happen in low-energy scattering processes, if there are at least two kinds of the interactions among the matter fields.

Consider a potential, which is composed of a Coulomb potential and a Yukawa potential
\begin{align}V(r) =V_C+V_Y
=- \frac{\alpha }{r}-\frac{{{\alpha _Y}}}{r}{e^{ - \mu r}}
\label{long-short}
\end{align}
\noindent
where  $V_C(r)$ is the Coulomb potential, $\alpha$ is its coupling
constant, and $V_Y(r)$ is the Yukawa potential, ${\alpha _Y}$ is the coupling constant, $\mu$ is the screening constant. This kind of potential can be produced by the lagrangian
\begin{equation}
\mathcal{L}_{int}=-\lambda\bar{\psi}\gamma^{\mu}\psi A_{\mu}-g\bar{\psi}\psi\phi
\end{equation}

It is well known that the resonance behavior is a significant feature of the Sommerfeld enhancement (see,eg.\cite{Lattanzi}), it corresponds to the zero-energy resonance scattering. In the beginning, let us study the resonance behavior of the multi-Sommerfeld enhancement.

For the potential
\begin{equation}
2mV(r)=-\frac{b}{4r^2}
\end{equation}
We know that if $b>1$ (in atomic units), there is an infinity of bound states produced by the potential, and if $b<1$, the number of the bound states of the potential is finite. It is easy to find that the potential (\ref{long-short}) has an infinity of bound states, because at large distances the potential decreases slower than the ``critical potential'' $2mV(r)=-1/4r^2$. An infinite number of bound states indicates that the bound states near zero-energy is quasi-continue in quasi-infinite space, thus there is no resonance in the scattering and the resonance behavior would absent in the Sommerfeld enhancement, as in the case of the Coulomb potential scattering. The naive argument indicates that the resonance behavior would also absent in the multi-Sommerfeld enhancement for the potential ({\ref{long-short}}). However, the numerical simulations of this kind of potential show that the resonance behavior could appear in the  multi-Sommerfeld enhancement, see figure 1.

\begin{figure}[h]
\begin{center}
{\includegraphics[width=6cm]{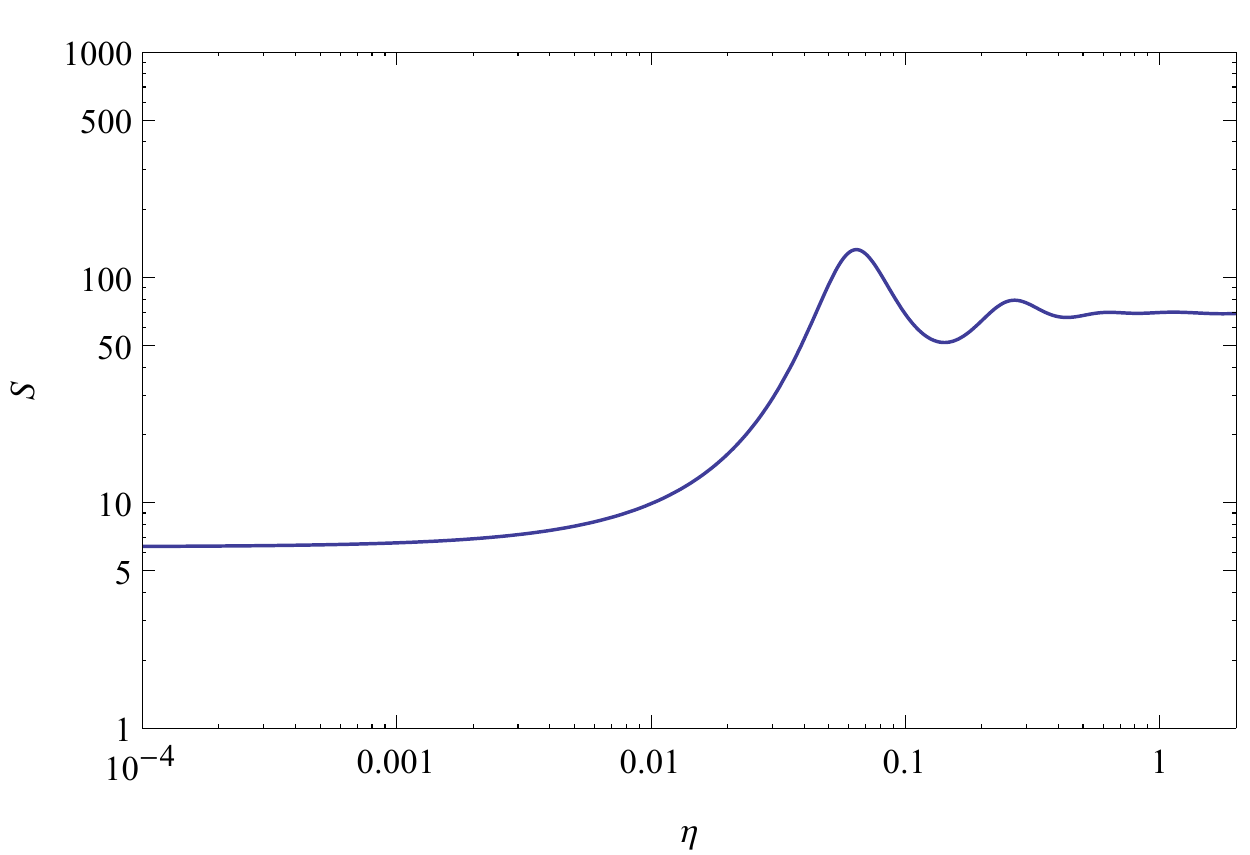}}
\caption{Multi-Sommerfeld factor for ${2\alpha/v}= 2$ and ${2\alpha_Y/v}= 20$, where $\eta=mv/\mu$. We can see that the resonance enhancement appears in this case.}
\end{center}
\end{figure}

To explain this resonance behavior of the multi-Sommerfeld enhancement, we need to study the special relation between the potential ({\ref{long-short}}) and bound states.

Notice that the Yukawa potential decreases exponentially and the Coulomb potential vanishes as $r^{-1}$ , thus at large distances
\begin{equation}
V(r)\longrightarrow -\frac{\alpha}{r}
\end{equation}
However, the situation becomes complicated at small distances.

If $\alpha_Y>\alpha$, near the origin, the Yukawa force is stronger than the Coulomb force. However, $V_Y(r)$ decreases much faster than $V_C(r)$, the Coulomb field would win out at large distances. To get the ``critical point'', we use the condition
\begin{equation}
\frac{\alpha}{r_0}=\frac{\alpha_Y}{r_0}e^{-\mu r_{0}}
\end{equation}
and find
\begin{equation}
r_0=\frac{1}{\mu}\ln\frac{\alpha_Y}{\alpha}
\end{equation}
It shows that there are two distinct regions of the potential ({\ref{long-short}}). In the region $r<r_0$, the Yukawa part dominates, and in the region $r>r_0$, the Coulomb part dominates. We may say that the potential ({\ref{long-short}}) decreases like the Yukawa potential at small distances but decreases like the Coulomb potential at large distances. Intuitively, if the Yukawa part of $V(r)$ is able to develop bound states, $V(r)$ could form shallow bound states in finite space. That explains why the resonance behavior could appear in the multi-Sommerfeld enhancement.

The situation changes if $\alpha_Y<\alpha$. The potential ({\ref{long-short}}) is dominated by the Coulomb part, and it decreases like the Coulomb potential at any distances. In this case, no matter the Yukawa part could or could not develop bound states, the potential (\ref{long-short}) is difficult to form shallow bound states in finite space, thus there is no resonance behavior in the multi-Sommerfeld enhancement, see figure 2.

\begin{figure}[h]
\begin{center}
{\includegraphics[width=6cm]{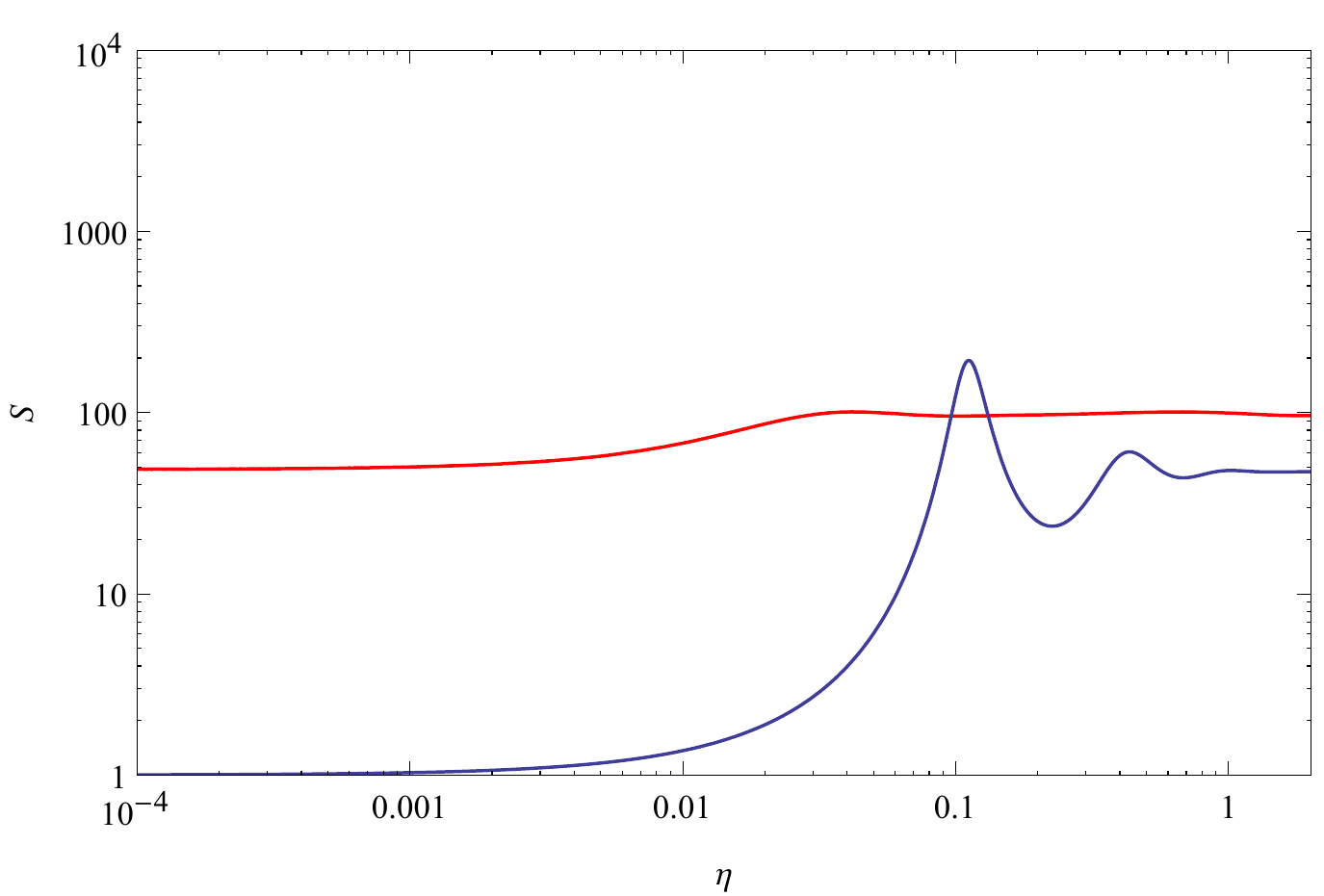}}
\caption{Multi-Sommerfeld factor (\textcolor{red}{red}) for $2\alpha/v=15$, $2\alpha_Y/v=15$, and the Sommerfeld factor (\textcolor{blue}{blue}) for the Yukawa potential, where $\eta=mv/\mu$.}
\end{center}
\end{figure}

As explained above, it is already shown that if $\alpha_Y>\alpha$, the bound states created by the Yukawa part could ``exist'' in the potential (\ref{long-short}). However, the bound states of the potential (\ref{long-short}) would not be exactly the same as the ones of the Yukawa part. To order to show its implication for the multi-Sommerfeld enhancement, we can define an effective Yukawa potential in the region $r\leq r_0$
\begin{equation}
V_{eff}(r)=-\frac{\alpha'}{r}e^{-\mu'r}
\end{equation}
where $\alpha'=\alpha+\alpha_Y$, and $\mu'$ is the effective screening constant. At the point $r=r_0$, it requires
\begin{equation}
\frac{\alpha_Y}{r_0}e^{-\mu r_0}+\frac{\alpha}{r_0}=\frac{\alpha'}{r_0}e^{-\mu\prime r_0}
\end{equation}
and we can get
\begin{equation}
\mu'=\mu\ln{\frac{2\alpha_Y}{\alpha+\alpha_Y}}
\end{equation}

It is well known that the ability of the Yukawa potential to develop bound states is proportional to the value of $m\alpha/\mu$, and exact value for existing a bound state of Yukawa potential is $2m\alpha/\mu=1.68$ \cite{Blatt}. A simple calculation shows
\begin{equation}
\frac{\alpha'}{u'}>\frac{\alpha_Y}{\mu}
\end{equation}
This result indicates that comparing with the single Yukawa potential case, the resonance positions of the multi-Sommerfeld enhancement for the potential (\ref{long-short}) would shift to the left side of $\eta$ (the side of smaller mass). This shift behavior of resonance positions is observed, see figure 3. The resonance enhancement would not be $1/v^2$-dependence as a consequence of the appearance of the long-range potential.
\begin{figure}[h]
\begin{center}
{\includegraphics[width=6cm]{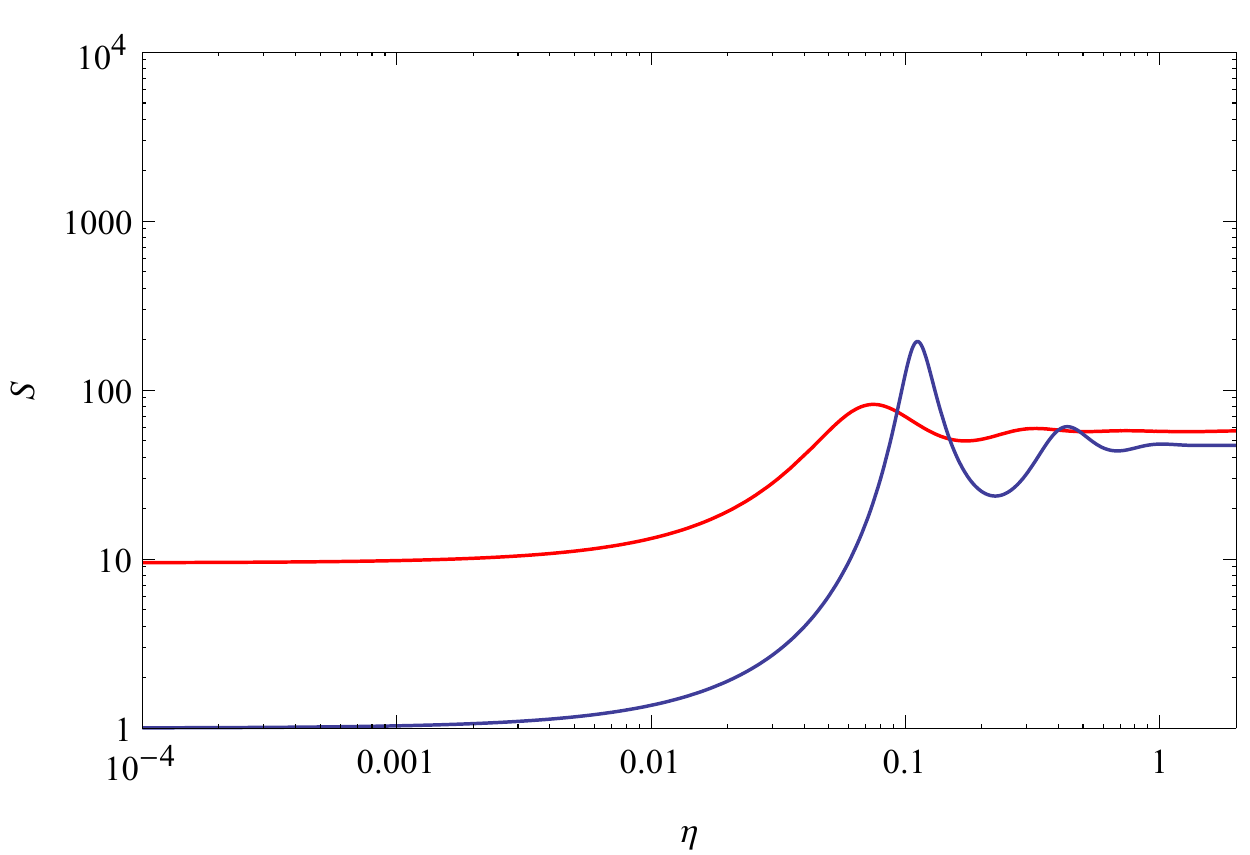}}
\caption{Multi-Sommerfeld factor (\textcolor{red}{red}) for $2\alpha/v=3$, $2\alpha_Y/v=15$, and the Sommerfeld factor (\textcolor{blue}{blue}) for the Yukawa part, where $\eta=mv/\mu$. We can see that the resonance positions do shift to the left as expected. The resonance enhancement is also alleviated by the Coulomb part.}
\end{center}
\end{figure}

As we know, in the case of the Sommerfeld enhancement for a Yukawa potential, the Sommerfeld factor is same as the one of the Coulomb potential, if $mv/\mu\gtrsim1$ (see, e.g.\cite{Arkani-Hamed}). We notice that this Coulomb approximation behavior still appears in the case of the multi-Sommerfeld enhancement, see figures 1-3.

So far we have ignored an interesting part of the multi-Sommerfeld mechanism: there is another stable region of the multi-Sommerfeld factor, where $\eta\ll v/2\alpha$ (i.e. $2m\alpha/\mu\ll1$). Although we could reasonably infer that the enhancement comes from the Coulomb part of the potential (\ref{long-short}), it would be more meaningful to us if we can show the formal aspect of this behavior.

Define the scattering operator
\begin{equation}
T_{fi}=\big{<}f|V_C+V_Y|\psi^{(+)}\big{>}
\label{two}
\end{equation}
where $\big{<}f|$ is the final plane wave state, and $|\psi^{(+)}\big{>}$ is defined by the Lippmann-Schwinger equation
\begin{equation}
|\psi^{(+)}\big{>}=|i\big{>}+\frac{1}{E-H_0+i\epsilon}(V_C+V_Y)|\psi^{(+)}\big{>}
\end{equation}
where $|i\big{>}$ is the initial plane wave state.

Recall that in the formal theory of the scattering, the Gell-Mann and Goldberger's transformation \cite{Gell-Mann} for the potential $U+V$ is
\begin{align}
A_{fi}&=\big{<}f|V+U|\psi^{(+)}\big{>}\nonumber\\
      &=\big{<}f^{(-)}|V|i^{(+)}\big{>}+\big{<}f|U||i^{(+)}\big{>}
\label{Gell-Mann}
\end{align}
where $|i^{(+)}\big{>}$ is defined by
\begin{equation}
|i^{(+)}\big{>}=|i\big{>}+\frac{1}{E-H_0+i\epsilon}U|i^{(+)}\big{>}
\end{equation}
and $|f^{(-)}\big{>}$ is defined by
\begin{equation}
|f^{(-)}\big{>}=|f\big{>}+\frac{1}{E-H_0-i\epsilon}U|f^{(-)}\big{>}
\end{equation}

We could introduce the incoming (outgoing) scattering state $|i^{(+)}_c\big>$ ($|f^{(-)}_c\big>$) for the non-perturbative Coulomb field. In order to simplify our discussion, the original Gell-Mann and Goldberger's transformation (\ref{Gell-Mann}) need be specially considered at the level of elementary particles: generally, the final states in the inelastic scattering would be extremely relativistic, $|f^{(-)}_c\big>$ can be simply replaced by $|f\big>$. We therefore could separate the scattering operator (\ref{two}) as
\begin{align}
T_{fi}&\equiv\big{<}f^{(-)}_c|V_C|i^{(+)}_c\big>+\big{<}f|V_Y|i^{(+)}_c\big>\nonumber\\
                                   &=\big{<}f|V_C|i^{(+)}_c\big>+\big{<}f|V_Y|i^{(+)}_c\big{>}
\label{Goldberger}
\end{align}
The first term in the right of Eq.(\ref{Goldberger}) would vanish if there are two massive scalar bosons in the final state, and the second term would vanish if there are two massless gauge bosons in the final state.

We know that $|i^{(+)}_c\big{>}$ behaves like \cite{Landau}
\begin{align}
\psi^{(+)}({\bf{r}})&=\big{<}{\bf{r}}|i^{(+)}_c\big>=B\big{<}{\bf{r}}|i\big{>}\nonumber\\
&=e^{\pi\alpha/2v}\Gamma(1-\frac{i\alpha}{v})\text{F}(\frac{i\alpha}{v},1,\frac{ivr-i{\bf v \cdot r}}{\alpha})e^{i{\bf k \cdot r}}
\end{align}
where F is the confluent hypergeometric function, defined by
\begin{equation}
\text{F}(\alpha,\gamma,z)=1+\frac{\alpha}{\gamma}\frac{z}{1!}+\frac{\alpha(\alpha+1)}{\gamma(\gamma+1)}\frac{z^2}{2!}+\cdots
\end{equation}
and
\begin{equation}
B=e^{\pi\alpha/2v}\Gamma(1-\frac{i\alpha}{v})\text{F}(\frac{i\alpha}{v},1,\frac{ivr-i{\bf v \cdot r}}{\alpha})
\end{equation}

If the interaction is point-like, the factor $B$ could be treated as a relevant constant of the scattering process, and we can formally redefine
\begin{equation}
T_{fi}=B\big{<}f|V_C|i\big{>}+B\big{<}f|V_Y|i\big{>}
\end{equation}

In our treatment, $B_{r=0}$ is the factor of the ``perturbation scattering amplitude'' for the interaction, and we obtain
\begin{equation}
S=\frac{{|\psi(0)|}^2}{|\psi_0|^2}=|B|^{2}_{r=0}=\frac{2\pi\alpha/v}{1-e^{-2\pi\alpha/v}}
\label{coulomb}
\end{equation}
We can see that this boost factor is exactly the multi-Sommerfeld factor in the region $\eta\ll v/2\alpha$.

Next let us consider another typical potential composed by the Yukawa potentials:
\begin{equation}
V(r) = V_1+V_2 = - \frac{{\alpha_1}}{r}e^{ - \mu_1 r}-\frac{{{\alpha_2}}}{r}{e^{ - \mu_2 r}}
\label{short-short}
\end{equation}
where $\alpha_i$ are the coupling constants, and ${\mu_i}$ are the screening constants, $i=1,2$. This kind of potential would appear in an abelian gauge theory, if the abelian gauge boson gets mass from the BEH mechanism.

For the potential (\ref{short-short}), the $s$-wave radical Schr\"{o}dinger equation is
\begin{equation}
\varphi ''(y) + (1 +\frac{1}{y{\varepsilon_1}}e^{ - \frac{y}{\eta_1 }} +\frac{1}{y{\varepsilon_2}}e^{ - \frac{y}{\eta_2 }})\varphi (y) = 0
\label{b-equation}
\end{equation}
where $\varepsilon_i=v/2\alpha_i$ and $\eta_i=mv/\mu_i$.

For different Yukawa potentials, the resonant Sommerfeld enhancements would have different resonance positions. We have already shown that the non-perturbative effects of the single potentials could be gradually turned on, it naturally suggest that there would be the multiple resonance behavior in the multi-Sommerfeld enhancement. For instance, we adopt $\varepsilon_1=10^{-1}=\varepsilon_2/5$, $\eta=\eta_1=10^{4}\eta_2$, and plot the boost factor, see figure 4.

\begin{figure}[h]
\begin{center}
{\includegraphics[width=6cm]{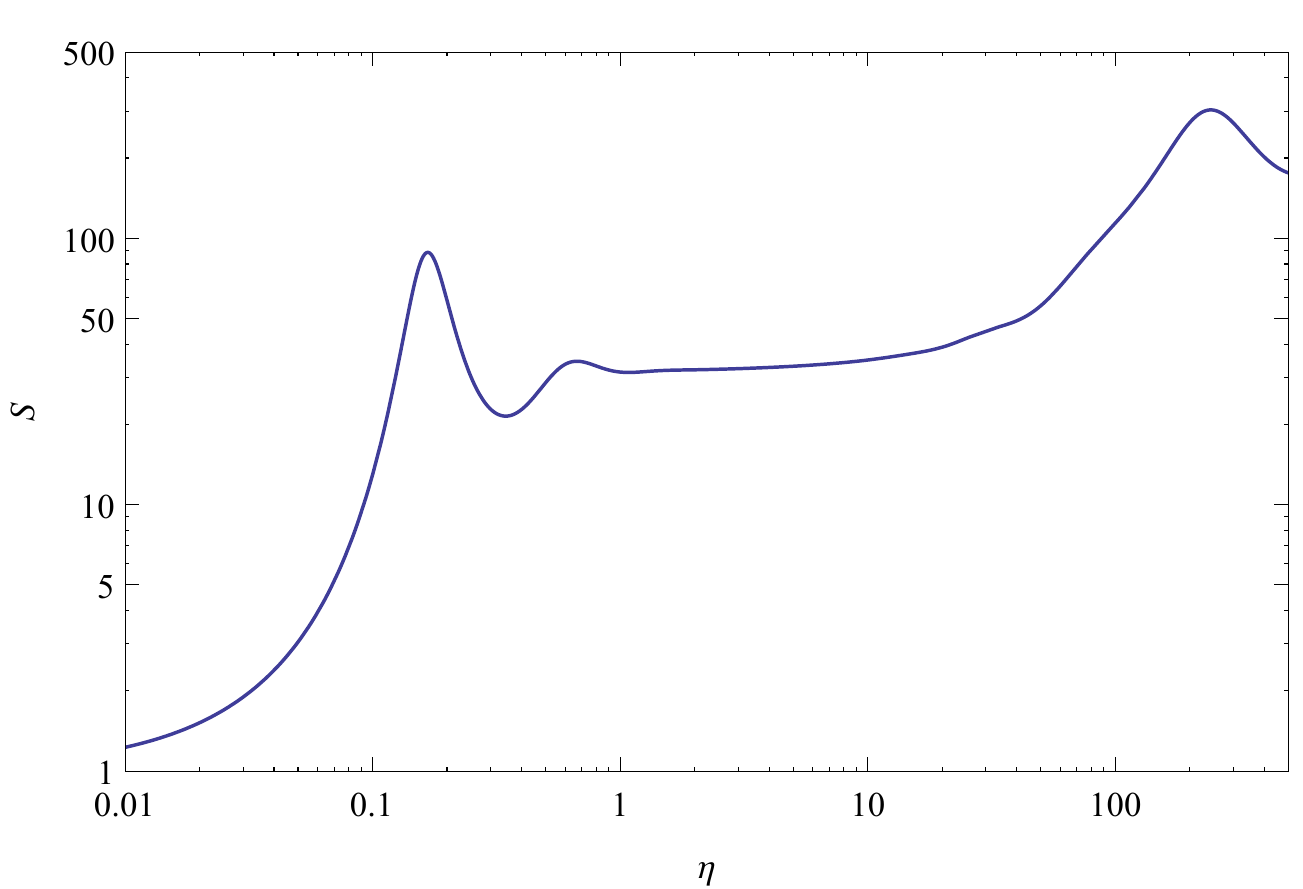}}
\caption{Multiple resonance behavior in the multi-Sommerfeld enhancement for $\varepsilon_1=v/2\alpha_1=0.1$, $\varepsilon_2=v/2\alpha_2=0.02$ and $\mu_1/\mu_2=0.0001$, where $\eta=\eta_1=mv/\mu_1$. }
\end{center}
\end{figure}
\noindent

If $V(r)$ is composed of a series of Yukawa potentials, we find that it is simple to generalize a conclusion of the single Yukawa potential case: for the multi-Sommerfeld enhancement, all the Yukawa potentials could be treated as the Coulomb potentials in the region $mv/\mu_{max}\gtrsim1$, where $\mu_{max}=\text{max}\{\mu_1,\mu_2,\ldots, \mu_n$\}.

\section{Phenomenology}

Recent years, the phenomenon of positron fraction excess in cosmic ray at high energy was observed by HEAT \cite{Beatty}, PAMELA \cite{Adriani}, Fermi-LAT \cite{Abdo} and AMS-02 \cite{Aguilar}. This anomaly could be explained by dark matter annihilation in WIMPs scenario, but if we want to keep ¡°WIMP miracle¡± surviving, large boost factors are needed. An attractive way of getting a large boost factor is to invoke the Sommerfeld enhancement \cite{Hisano}.

It was discussed in many models that dark matter could interact with itself via different kinds of gauge bosons (see, eg.(\cite{Hisano3,Cirelli2,Fargion}). For TeV-scale dark matter, it was suggested that there is a new GeV-scale force carrier $\phi$ in dark sector \cite{Arkani-Hamed}. This light force carrier would produce the significant Sommerfeld enhancement factor and make the dark matter annihilates dominantly into leptons.

However, if there is an unbroken $U(1)$ symmetry in dark sector, besides the light force carrier, the dark photon appears\footnote{Existing an unbroken $U(1)$ symmetry in dark sector is not a new idea, see. e.g.\cite{Fargion} or a comprehensive work \cite{Ackerman} and its references.}. Now TeV-scale dark matter could interact through a pure attractive potential:
\begin{equation}
V(r)= -\frac{{{\alpha _\phi}}}{r}{e^{ -{m_\phi }r}} -\frac{\alpha_D}{r}
\label{multi}
\end{equation}
\noindent
where ${m_\phi }$ is the mass of the light force carrier, $\alpha_\phi$ is its coupling constant, and $\alpha_D$ is the dark Coulomb coupling constant. Large dark Coulomb coupling constants are not favored by generally astrophysical observations (at least for simple models). Typically, we assume ${\alpha_D}<{\alpha _\phi}\thicksim10^{-2}$.

In order to show the implication of the multi-Sommerfeld enhancement for potential (\ref{multi}), let us consider the boost in substructure. The importance of the substructure boost was studied in literature \cite{Arkani-Hamed,Bovy,Kamionkowski,Slatyer2,Abazajian}. The subhalos have velocity dispersions much smaller than the $150$km/s of the smooth halo, it could be assumed that the Sommerfeld enhancement in subhalos saturates at $\thicksim m\alpha_{\phi}/m_{\phi}$. However, if dark photon also appears, the dark Coulomb part of the multi-Sommerfeld enhancement would not saturate at any velocities. We know that dwarf galaxies have the velocity dispersions of $\thicksim10^{-5}$ \cite{Strigari}. Thus the dark Coulomb interaction would dominate the multi-Sommerfeld enhancement in dwarf galaxies, if the dark Coulomb coupling constant satisfies
\begin{equation}
10^5\pi\alpha_D\gtrsim \frac{m\alpha_\phi}{m_\phi}
\label{condition}
\end{equation}
In order to give a quantitative result, we adopt $m_\phi=0.5$GeV, $\alpha_\phi=0.01$, $m=1$TeV. Using the Eq.(\ref{condition}), we can see that the boost in dwarf galaxies mainly comes from the dark Coulomb interaction, if $\alpha_D\gtrsim10^{-4}$. It shows that although the dark Coulomb coupling is much weaker than the coupling of the GeV-scale boson, the dark Coulomb interaction could dominate the multi-Sommerfeld enhancement in substructure. Generally, such small value of $\alpha_D$($\thicksim10^{-4}$) is potentially safe in current astrophysics, and appropriate value of $\alpha_\phi$ could produce the correct relic abundance.

\section{Summary}

Instead of investigating the comprehensive implication of the multi-Sommerfeld enhancement for TeV-scale dark matter, in this work, we concentrate on the physical content of the multi-Sommerfeld mechanism.

In the case where $V(r)$ is composed of a Coulomb potential and a Yukawa potential, we find that if the coulomb coupling constant is larger than the Yukawa coupling constant, $V(r)$ can not form shallow bound states in finite space. Thus there is no resonance behavior in the multi-Sommerfeld enhancement. If the coulomb coupling constant is smaller than the Yukawa coupling constant, there would be resonance behavior in the enhancement. Comparing with the Yukawa case, detail calculations show that the resonance positions would shift to smaller values of $m$. By means of a formal method, we prove that the enhancement in the range $2m\alpha/\mu\ll1$ is produced by the Coulomb part of the potential.

In the case where $V(r)$ is composed of a series of Yukawa potentials, the multiple resonance behavior of the multi-Sommerfeld enhancement is found. In the parameter space $mv/\mu_{max}\gtrsim1$, the Yukawa parts in the potential could be approximated as the Coulomb potentials, and the Coulomb approximation behavior of the multi-Sommerfeld enhancement appears.

Although we focus on the multi-Sommerfeld mechanism itself, its potential influence on dark sector is briefly discussed. We show that for an unbroken $U_D(1)$ symmetry in dark matter, even the dark Coulomb coupling constant is two orders of magnitude smaller than the coupling constant of the GeV-scale force carrier, attributing to the small velocities of dark matter subhalos and the saturation behavior of the short-range potential, the dark Coulomb interaction could dominate the boost in dwarf galaxies.

\appendix

\section{Sommerfeld enhancement for different parameter spaces}
In this appendix, we show how to get different parameter spaces of the (multi-)Sommerfeld enhancement.

The $s$-wave radical Schr\"{o}dinger equation for the Yukawa potential is
\begin{equation}
\frac{1}{{2m}}\frac{{{d^2}\phi(r)}}{{d{r^2}}} + \frac{{{\alpha _Y}}}{r}{e^{ - \mu r}} \phi(r) =  - \frac{{m{v^2}}}{2}\phi(r)
\label{schroedinger}
\end{equation}
\noindent
where $\phi(r) = r{R_{l = 0}}(r)$.

To obtain the Sommerfeld factor, we need solve it numerically with boundary
conditions:
\begin{equation}\phi(0)=0
~~\text{and}~~
 \phi(r) \to 2\sin({kr + \delta })\text{ as $r \to \infty$ }
\end{equation}

It is convenient to introduce scaled variable $x=2{\alpha _Y}mr$, and we can rewrite the Eq.(\ref{schroedinger}) as
\begin{equation}
\phi ''(x) + (\varepsilon _Y^2 + \frac{1}{x}{e^{-{\varepsilon _\mu}x}})\phi (x) = 0
\label{equation1}
\end{equation}
where ${\varepsilon _{_Y}} = v/2{\alpha _Y}$, ${\varepsilon _\mu } = \mu /2{\alpha _Y}m$. The Sommerfeld enhancement depends on ${\varepsilon _Y}$ and ${\varepsilon_\mu}$.

As we know, in the case of the Coulomb scattering, we can redefine variables to get the standard Kummer's equation from the Whittaker's equation, and it does not alter the physically observable quantities. Similarly, we can redefine $y = vx/2\alpha_Y  = kr$ and $\varphi  = k\phi$ \cite{Iengo}, the Eq.(\ref{equation1}) becomes
\begin{equation}
\varphi ''(y) + (1 + \frac{1}{{y{\varepsilon _Y}}}{e^{ - \frac{y}{\eta }}})\varphi (y) = 0
\label{equation2}
\end{equation}
\noindent
where $\eta=mv/\mu$. Now, the Sommerfeld enhancement depends on $\eta$ and ${\varepsilon _Y}$.

For the potential (\ref{long-short}), we get
\begin{equation}
\varphi ''(y) + (1 +\frac{1}{y\rho} +\frac{1}{{y{\varepsilon _Y}}}{e^{ - \frac{y}{\eta }}})\varphi (y) = 0
\label{a-equation}
\end{equation}
where $\rho=v/2\alpha$, and the multi-Sommerfeld enhancement depends on $\eta$, ${\varepsilon _Y}$ and $\rho$.


\begin{thebibliography}{99}
\bibitem{Sommerfeld}
  A.~Sommerfeld,
  Annalen\ der\ Physik {\bf 403}, 257 (1931).

\bibitem{Sakharov}
  A.~D.~Sakharov,
  Zh. Eksp. Teor. Fiz. {\bf18}, 631 (1948) [Sov.\ Phys.\ Usp. {\bf34}, 375 (1991)].

\bibitem{Hisano}
  J.~Hisano, S.~Matsumoto, M.~M.~Nojiri and O.~Saito,
  Phys.\ Rev.\ D {\bf 71}, 063528 (2005).

\bibitem{Hisano2}
  J.~Hisano, S.~Matsumoto, M.~Nagai, O.~Saito and M.~Senami,
  Phys. Lett. B {\bf646} 34 (2007).

\bibitem{Cirelli}
  M.~Cirelli, A.~Strumia and M.~Tamburini,
  Nucl.\ Phys.\ B {\bf800}, 204 (2008).

\bibitem{March-Russell}
  J.~March-Russell, S.~M.~West, D.~Cumberbatch and D.~Hooper,
  JHEP {\bf0807}, 058 (2008).

\bibitem{Arkani-Hamed}
  N.~Arkani-Hamed, D.~P.~Finkbeiner, T.~R.~Slatyer and N.~Weiner,
  Phys.\ Rev.\ D {\bf 79}, 015014 (2009).

\bibitem{Lattanzi}
  M.~Lattanzi and J.~Silk,
  Phys.\ Rev.\ D {\bf 79}, 083523 (2009).

\bibitem{March-Russell2}
  J.~D.~March-Russell and S.~M.~West,
  Phys.\ Lett.\ B {\bf676} 133 (2009).

\bibitem{Strumia}
  A.~Strumia,
  Nucl.\ Phys.\ B {\bf809}, 308 (2009).

\bibitem{Iengo}
  R.~Iengo,
  JHEP {\bf 0905}, 024 (2009).

\bibitem{Cassel}
  S.~Cassel,
  J.\ Phys.\ G {\bf 37}, 105009 (2010).

\bibitem{Slatyer}
  T.~R.~Slatyer,
  JCAP {\bf 1002}, 028 (2010).

\bibitem{Blatt}
  J.~M.~Blatt and J.~D.~Jackson,
  Phys.\ Rev. {\bf76}, 18 (1949).

\bibitem{Gell-Mann}
  M.~Gell-Mann and M.~L.~Goldberger,
  Phys.\ Rev.\ {\bf91}, 398 (1953).

\bibitem{Landau}
  L.~D.~Landau and E.M.Lifshitz,
  \textsl{Quantum Mechanics}, Pergamon Press 1958.

\bibitem{Beatty}
  J.~J.~Beatty $et\ al$.,
  Phys.\ Rev.\ Lett. {\bf 93}, 241102 (2004).

\bibitem{Adriani}
  O.~Adriani $et\ al$.,
  Nature {\bf 458}, 607 (2009).

\bibitem{Abdo}
  A.~A.~Abdo $et\ al$.,
  Phys.\ Rev.\ Lett. {\bf 102}, 181101 (2009).

\bibitem{Aguilar}
  M.~Aguilar $et\ al$.,
  Phys.\ Rev.\ Lett. {\bf 110}, 141102 (2013).

\bibitem{Hisano3}
  J.~Hisano, S.~Matsumoto and M.~M.~Nojiri,
  Phys.\ Rev.\ D {\bf67}, 075014 (2003).

\bibitem{Cirelli2}
  M.~Cirelli, A.~Strumia and M.~Tamburini,
  Nucl.\ Phys.\ B {\bf787}, 152 (2007).

\bibitem{Fargion}
  D.~Fargion,~M.~Y.~Khlopov and C.~A.~ Stephan,
  Class.\ Quantum.\ Grav. {\bf23}, 7305 (2006).

\bibitem{Ackerman}
  L.~Ackerman, M.~R.~Buckley, S.~M.~Carroll and M.~Kamionkowski,
  Phys.\ Rev.\ D {\bf 79}, 023519 (2009).

\bibitem{Bovy}
  J.~Bovy,
  Phys.\ Rev.\ D {\bf79}, 083539 (2009).

\bibitem{Kamionkowski}
  M,~Kamionkowski, S.~M.~Koushiappas and M.~Kuhlen,
  Phys.\ Rev.\ D {\bf81}, 043532 (2010).

\bibitem{Slatyer2}
  T.~R.~Slatyer, N.~Toro and N.~Weiner,
  Phys.\ Rev.\ D {\bf86}, 083534 (2012).

\bibitem{Abazajian}
  K.~N.~Abazajian and J.~P.~Harding,
  JCAP {\bf1201}, 041 (2012).

\bibitem{Strigari}
  L.~E.~Strigari, S.~M.~Koushiappas, J.~S.~Bullock and M.~Kaplinghat,
  Phys.\ Rev.\ D {\bf75}, 083526 (2007).


\end{thebibliography}
\end{document}